\documentclass[twocolumn,aps,amsmath,amssymb,prl]{revtex4}
\usepackage{epsfig,bm,dcolumn}
\usepackage{graphicx}
\usepackage{epsfig}
\usepackage{latexsym}
\usepackage{bm}

\begin{document}

\renewcommand{\ni}{{\noindent}}
\newcommand{\dprime}{{\prime\prime}}
\newcommand{\be}{\begin{equation}}
\newcommand{\ee}{\end{equation}}
\newcommand{\bea}{\begin{eqnarray}} 
\newcommand{\eea}{\end{eqnarray}}
\newcommand{\nn}{\nonumber} 
\newcommand{\bk}{{\bf k}}
\newcommand{\bQ}{{\bf Q}}
\newcommand{\q}{{\bf q}}
\newcommand{\s}{{\bf s}}
\newcommand{\bN}{{\bf \nabla}}
\newcommand{\bA}{{\bf A}}
\newcommand{\bE}{{\bf E}}
\newcommand{\bj}{{\bf j}}
\newcommand{\bJ}{{\bf J}}
\newcommand{\bs}{{\bf v}_s}
\newcommand{\bn}{{\bf v}_n}
\newcommand{\bv}{{\bf v}} 
\newcommand{\la}{\langle}
\newcommand{\ra}{\rangle} 
\newcommand{\dg}{\dagger}
\newcommand{\br}{{\bf{r}}} 
\newcommand{\brp}{{\bf{r}^\prime}} 
\newcommand{\bq}{{\bf{q}}}
\newcommand{\hx}{\hat{\bf x}} 
\newcommand{\hy}{\hat{\bf y}}
\newcommand{\bS}{{\bf S}} 
\newcommand{\cU}{{\cal U}}
\newcommand{\cD}{{\cal D}} 
\newcommand{\bR}{{\bf R}}
\newcommand{\pll}{\parallel} 
\newcommand{\sumr}{\sum_{\vr}} 
\newcommand{\cP}{{\cal P}} 
\newcommand{\cQ}{{\cal Q}} 
\newcommand{\cS}{{\cal S}}
\newcommand{\ua}{\uparrow} 
\newcommand{\da}{\downarrow}

\title{Strong correlations lead to protected low energy excitations in disordered d-wave superconductors}

\vspace{-1.0cm}
\author{Arti Garg$^{1}$, Mohit Randeria$^{2}$ and Nandini Trivedi$^{2}$}

\affiliation{$^{1}$ Department of Theoretical Physics, Tata Institute of
Fundamental Research, Mumbai 400005, and Department of Physics, Indian Institute of Science, Bangalore, India\\
$^{2}$ Department of Physics, The Ohio State University, Physics Research Building, 191 W. Woodruff Avenue,
Columbus, OH 43210}

\begin{abstract}
\vspace{0.4cm}
\noindent
We show that strong correlations play a vital role in protecting low energy excitations in disordered high temperature superconductors. 
The impurity-induced low-energy density of states (DOS) is greatly reduced in the strongly correlated superconductor compared to d-wave Bogoliubov-deGennes theory.
The gapless nodal quasiparticles, and the resulting `V' in the low-energy DOS, are much more robust against disorder compared to the large-gap antinodal excitations. 
We discuss the relevance of our results to angle-resolved photoemission and scanning tunneling spectroscopy experiments.
\end{abstract}

\maketitle

\noindent
The effect of disorder on the strongly correlated d-wave superconducting state of high $T_c$ cuprates is a 
problem of great interest because disorder is \emph{intrinsic} to doped Mott
insulators. The resulting nanoscale inhomogeneity and its effects on the properties of cuprates has been investigated
in a series of beautiful scanning tunneling spectroscopy (STS) experiments \cite{pan,mcelroy,yazdani}.

One of the earliest puzzles regarding impurity effects~\cite{balatsky,pjh} in the cuprates was: 
\emph{why} is d-wave superconductivity (SC) so robust against disorder?
It was realized over the years that there are at least two reasons for this: (i) the intrinsic disorder lies
off the CuO$_2$ planes in the spacer layers, and (ii) even in the case of intentional disorder, such as Zn on the planar
Cu sites, the short coherence length in the cuprates leads to a very local and inhomogeneous response to impurities
\cite{uemura,ghosal} in marked contrast to standard Abrikosov-Gorkov theory.

In this paper we show that there is another important reason: 
\emph{disorder effects are greatly suppressed in the presence of strong correlations,} which are of course
central to high Tc superconductivity. To make this point, we compare the results of two different $T=0$
calculations. (A) The first includes
disorder-induced inhomogeneity in the superconductor using the Bogoliubov-deGennes (BdG) framework, but
ignores correlation effects. In fact
most of the theoretical work ~\cite{balatsky,pjh,ghosal,nunner} in the field, barring
a few exceptions \cite{z.wang}, uses just such an approach. 
(B) The second calculation includes the effects of both
disorder and correlations, combining the inhomogeneous BdG approach 
with the projection to states with no-double occupancy, as explained in 
greater detail below. By comparing the results of (A) and (B) we find that:

\noindent
(1) Strong correlations lead to a renormalization of impurity effects, resulting in a
much shorter healing length over which the pairing amplitude is destroyed in the vicinity of an impurity.

\noindent 
(2) The same impurity potential generates significantly fewer low energy excitations in the 
correlated system.

\noindent 
(3) Gapless nodal quasiparticles are much less affected by disorder compared with the large-gap excitations 
near the Brillouin zone edge. 

\noindent 
(4) The robustness of the nodal quasiparticles can be seen by the very small impact of impurities on the ``V'' in
the density of states $N(\omega)$, i.e., on its $|\omega|$ behavior at low energy, in the strongly correlated system.

\noindent 
(5) The point nodes are protected in the correlated system in the presence of disorder. In contrast, in the absence of correlations, the
point nodes of the clean system expand
into extended arc-like regions of gapless excitations induced by disorder.

The robustness of the nodes and nodal quasiparticles as well as the nodal-antinodal dichotomy
are of direct relevance to recent angle-resolved photoemission spectroscopy (ARPES)\cite{mohit_arpes,zhou,shen} 
and STS experiments ~\cite{mcelroy,yazdani}.

\bigskip

\noindent {\bf Model and Method:}
We use the t-t$^{\prime}$-J model
\bea
 H = -\sum_{{{\bf r},{\bf r}^\prime},\sigma}t_{{\bf r}{\bf r}^\prime}(c^{\dagger}_{{\bf r}\sigma}c_{{{\bf r}^\prime}\sigma}+h.c.)
 \hspace{3cm}\nonumber \\
 + J\sum_{<{{\bf r},{\bf r}^\prime}>}({\bf S}_{\bf r}\stackrel{.}{}{\bf S}_{{\bf r}^\prime}-n_{\bf r}n_{{\bf r}^\prime}/4) 
 + \sum_{\bf r}(V({\bf r})-\mu)n_{\bf r}     
\label{tJH}
\eea
with a random potential $V({\bf r})$ to describe the effects of disorder in a strongly correlated superconductor. 
$H$ acts on a Hilbert space with no doubly-occupied sites, resulting from the on-site Coulomb $U$,
the largest energy scale in the problem. The kinetic energy describes electrons with spin 
$\sigma$ hopping from site ${\bf r}$ to ${\bf r}^\prime$ on a 2D square lattice, with 
$t$ the near-neighbor and $t^\prime$ the next-near-neighbor hopping amplitudes. 
The superexchange $J=4t^2/U$ is responsible for d-wave pairing in the doped system and 
$\mu$ fixes the average density $n = 1 - x$ with hole doping $x$ ~\cite{noteU0}. 
The impurity potential $V({\bf r})$ is $V_{0}>0$ at randomly located sites with a concentration $n_{imp}$ 
and is zero otherwise. We will focus on weak potentials $V_0 = t$ as a simple model of intrinsic 
disorder in cuprates.

The most commonly used approximation is to simply ignore the no-double-occupancy constraint
and solve the BdG mean field equations for an inhomogeneous SC \cite{pjh,ghosal}. 
The interaction in eq.~(\ref{tJH}) is decomposed 
into {\em local} bond pairing amplitudes $\Delta_{{{\bf r}{\bf r}^\prime}} = 
J\langle c_{{\bf r}\uparrow}c_{{{\bf r}^\prime}\downarrow} - c_{{\bf r}\downarrow}c_{{{\bf r}^\prime}\uparrow}\rangle$/2, and 
Fock shifts $W_{{{\bf r}{\bf r}^\prime}} =J\langle c^{\dagger}_{{{\bf r}^\prime}\sigma}c_{{\bf r}\sigma}\rangle/2 $ with the density 
$n({\bf r}) =\langle \sum_{\sigma}c^{\dagger}_{{{\bf r}}\sigma}c_{{\bf r}\sigma}\rangle$. The effective BdG Hamiltonian is then 
diagonalized and the fields defined above calculated self-consistently. 
The results obtained by ignoring correlations will be called `BdG results' from now on. 

Our main goal is to include strong correlations over and above the inhomogeneous BdG theory.
The Hilbert space consists of states $|\Phi\rangle = P|\Phi_0\rangle$ where the projection operator
$P=\prod_{{\bf r}}(1-n_{{\bf r}\ua}n_{{\bf r}\da})$ suppresses 
double occupancy in any state $|\Phi_0\rangle$. We use here the Gutzwiller renormalized mean field theory \cite{zhang,vanilla} to handle 
projection. In the translationally invariant, clean system such an approximate treatment is in very 
good semi-quantitative agreement \cite{vanilla} with numerical studies of variational wavefunctions with $P$ implemented 
exactly \cite{paramekanti}. 

We next generalize the Gutzwiller approximation to inhomogeneous states. 
The main idea is to write the expectation value of any operator ${\cal Q}$ in
a state $P|\Phi_0\rangle$ as the product of a Gutzwiller factor $g_Q$ times the 
expectation value in $|\Phi_0\rangle$ so that 
$\langle {\cal Q} \rangle \simeq g_Q \langle {\cal Q} \rangle_0$.
The standard procedure \cite{zhang} for calculating $g_Q$ can be generalized
to keep track of the local density $x({\bf r}) = 1 - n({\bf r})$.
We thus obtain 
the kinetic energy $\langle c^{\dagger}_{{\bf r}\sigma}c_{{{\bf r}^\prime}\sigma}\rangle \approx 
g_t({{\bf r},{\bf r}^\prime}) \langle c^{\dagger}_{{\bf r}\sigma}c_{{{\bf r}^\prime}\sigma}\rangle_{0}$ 
and the spin correlation $\langle {\bf S}_{{\bf r}}\cdot{\bf S}_{{\bf r}^\prime}\rangle \approx  
g_s({{\bf r},{\bf r}^\prime}) \langle {\bf S}_{{\bf r}}\cdot{\bf S}_{{\bf r}^\prime}\rangle_{0}$.
We find the \emph{local} Gutzwiller factors are given by
$g_t({{\bf r},{\bf r}^\prime})={g_t({\bf r})g_t({\bf r}^\prime})$ with
$g_t({\bf r})=[2x({\bf r})/(1+x({\bf r}))]^{1/2}$
and 
$g_s({{\bf r},{\bf r}^\prime})=4/[(1+x({\bf r}))(1+x({\bf r}^\prime))]$.

\begin{figure}[h!]
\begin{center}
\includegraphics[width=1.8in,height=1.8in,angle=0]{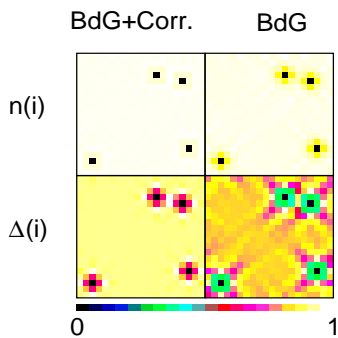}
\caption
{
Local density $n({\bf r})$ and pairing amplitude $\Delta({\bf r})$ variations, 
plotted on a \emph{normalized} scale of zero to one,
with $V_0=1.0t$, $n_{imp}=0.01$ and average $n=0.8$. 
Left panels: `BdG plus correlations' results  ($J=0.33t$). Right panels: 
`BdG' results ($J=1.1t$; see text).  
}
\label{fig1}
\end{center}
\vskip-6mm
\end{figure}

The BdG equations are then solved \cite{periodic_repetition} for eigenvalues $E_n$
eigenvectors $\left( u_n({\bf r}), v_n({\bf r})\right)$
together with \emph{local} self-consistency equations for
the density 
$n({\bf r}) = 2 \sum_{n} | v_{n}({\bf r})|^{2}$,
the pairing amplitude
$\Delta_{{\bf r, r^\prime}} = J_1({\bf r, r^\prime}) 
\sum_{n}[u_{n}({\bf r^\prime})v_{n}^{\star}({\bf r})+u_{n}({\bf r})v_{n}^{\star}({\bf r^\prime})]$
and the Fock shift
$W_{{\bf r, r^\prime}} = J_2({\bf r, r^\prime})\sum_{n}v_{n}({\bf r^\prime})v_{n}^{\star}({\bf r})$,
where $J_{1,2}({\bf r, r^\prime})=J(3g_s({{\bf r},{\bf r}^\prime}) \pm 1)/4$.

Results which include correlation effects will be denoted by `BdG plus Correlations' below. 
To meaningfully compare results with and without correlations, we choose all 
parameters, except $J$, to be same: $t=1$, $t^{\prime}=-0.25t$, impurity potential $V_0= t$, and work at 
the same values of average density $n$ and impurity concentration $n_{imp}$.
$J$ values are chosen such that we get the same $\Delta$ \emph{in the absence of disorder}  
in the two calculations.

\begin{figure}[h!]
\begin{center}
\includegraphics[width=3.0in,angle=0]{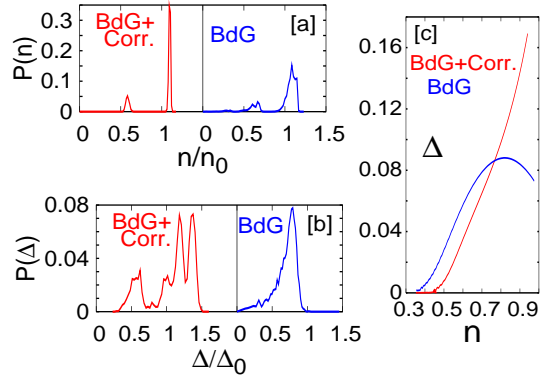}
\caption{Probability distributions (a) $P(n)$ of density $n({\bf r})$, and (b) $P(\Delta)$
of pairing $\Delta({\bf r})$ averaged over 20 disorder configurations for $n_{imp}=0.2$
(other parameters same as in Fig.~1). $n_0= 0.8$ and $\Delta_0 = 0.086$ correspond to the
case with no disorder ($n_{imp}=0$). (c) $\Delta(n)$ with and without correlations for $n_{imp}=0$,
with $J$ values chosen so that the corresponding $\Delta$ values match at $n = 0.8$.
}
\label{prob}
\end{center}
\vskip-6mm
\end{figure}

\bigskip

\noindent {\bf Local density and pairing amplitude:}
In Fig.~\ref{fig1} we compare the results obtained with (left panels) and without (right panels) correlation effects. In both cases 
disorder makes the density inhomogeneous, with $n({\bf r})$ suppressed near the repulsive impurity. There is a
marked difference in the spatial variation of the d-wave pairing amplitude 
$\Delta({\bf r})=\sum_{\bf r^\prime} \varepsilon_{{\bf r, r^\prime}} \Delta_{{\bf r, r^\prime}}/4$ where
$\varepsilon_{{\bf r, r^\prime}} = 1$ for ${\bf r^\prime} = {\bf r} \pm \hat{\bf x}$ and is $-1$ for
${\bf r^\prime} = {\bf r} \pm \hat{\bf y}$. In the BdG calculations the impurity affects 
$\Delta({\bf r})$ over a longer distance compared with the highly local response in the correlated system. 
One might think that this difference in healing lengths comes from the factor of two mass renormalization
\cite{paramekanti} due to correlations observed in the non-disordered case, however, we have checked 
\cite{check} that in fact the spatially varying Gutzwiller factors are responsible for the observed effect in Fig.~1. 
The suppressed electron density near an impurity modifies the local Gutzwiller factors and 
leads to an enhanced hopping on the bonds around it.
The system is consequently able to repair the damage induced by the disorder and heal the pairing amplitude over a shorter length scale.

We plot in Fig~\ref{prob}(a) and (b) the normalized distributions $P(n)$ of the density $n({\bf r})$, and $P(\Delta)$
of the pairing $\Delta({\bf r})$. $P(n)$, which is a delta-function at $n_0=0.8$ for the pure system, develops 
a bimodal structure in the disordered case. The low $n$ peak comes from sites at or near the impurities, 
while the $n>n_0$ peak leads to an average density $n_0$.
We see that correlations produce a sharper $P(n)$ compared to the simple BdG calculation. 
In the Bdg calculation $P(\Delta)$ is broadened over a range of $\Delta$ values strictly below  
the zero disorder $\Delta_0$. In contrast, in the correlated system, $P(\Delta)$ has weight both above and
below $\Delta_0$. This difference in $P(\Delta)$ with and without correlations can be understood from $\Delta(n)$
variation in the uniform, clean system (Fig.~\ref{prob}(c)). 
The structure in $P(\Delta)$ in the correlated system can also be understood in detail, with various
peaks associated with different kinds of sites: at, near and far from an impurity.

\vspace {-3cm}
\begin{figure}[h!]
\begin{center}
\hspace*{-0.5cm}
\includegraphics[scale=0.4,angle=-90]{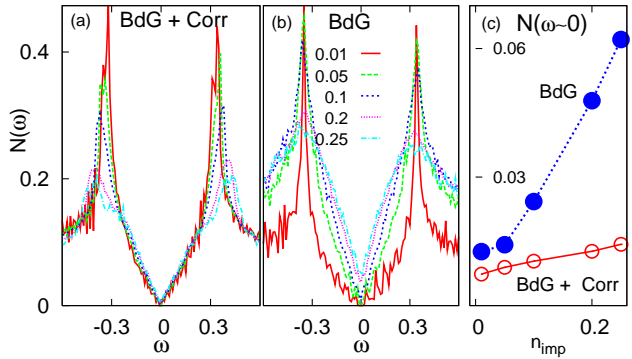}
\caption{
DOS $N(\omega)$ for various $n_{imp}$ [listed in panel (b)] averaged over 15 impurity configurations. Panel (a) shows results
with and (b) without correlation effects. (Other parameters are the same as Fig.~1). The energy gap is $4\Delta_0$ in the pure limit.
(c) Low energy DOS integrated over a small window $|\omega| \le 0.02t$ as a function of $n_{imp}$.
Results with correlations shown with open symbols (red) and 
plain BdG with closed symbols (blue).
}
\label{dos}
\end{center}
\vskip-6mm
\end{figure}

\bigskip

\noindent {\bf Tunneling density of states:}
In a disordered system the one-particle spectral function 
$A({\bf r},{\bf R};\omega) = - {\rm Im}G({\bf r}_1,{\bf r}_2;\omega+ i0^+)/\pi$ 
depends on both the center-of-mass ${\bf R} = ({\bf r}_1 + {\bf r}_2)/2$ and the relative 
coordinate ${\bf r} = {\bf r}_1-{\bf r}_2$. Within the Gutzwiller approximation we obtain 
$A({\bf r},{\bf R};\omega) = g_t({\bf r}_1,{\bf r}_2)A^0({\bf r},{\bf R};\omega)$ with
$A^0 = \sum_n u_n({\bf r}_1)u_n^{\star}({\bf r}_2)\delta(\omega-E_n)+
v_n({\bf r}_1)v_n^{\star}({\bf r}_2)\delta(\omega+E_n)$.
The Gutzwiller approximation describes the coherent part of the spectral function \cite{sumrules}
which dominates $A$ for $\omega$'s smaller than or comparable to the gap.
We focus exclusively on low-energy excitations here~\cite{incoherent}.

STM experiments measure the local density of states (DOS) 
$N({\bf R},\omega)=A(0,{\bf R};\omega)$.
We calculate the total DOS $N(\omega)$ obtained by averaging $\langle N({\bf R},\omega) \rangle_{{\bf R}}$
over the system and find that there are very significant differences between the results with 
(Fig.~\ref{dos}(a)) and without correlations (Fig.~\ref{dos}(b)). 
The most striking observation is that the $|\omega|$ behavior in the low energy DOS in panel (a) 
is hardly affected as $n_{imp}$ increases from 1\% to 25\%, in marked contrast with panel (b). Further, 
by integrating $N(\omega)$ over a small window $|\omega| \le 0.02t$, we see in
Fig.~\ref{dos}(c) that very little low energy spectral weight is generated in the correlated system from
the pair breaking effect of impurities compared with plain BdG calculation. 
This leads to one of our central results: low energy excitations are much more robust against disorder
in the correlated system compared with the simple BdG calculation. 

Next we look at the DOS near the gap edge. In both Figs.~\ref{dos}(a) and (b) the sharp log singularities (``coherence peaks'') 
seen in the clean d-wave SC are suppressed with increasing $n_{imp}$. 
The energy scale of the peaks in the DOS is shifted up
slightly with increasing $n_{imp}$ in the correlated case (see Fig.~\ref{dos}(b)), while it seems more or less constant in the simple BdG results.

\begin{figure}[h!]
\begin{center}
\includegraphics[width=2.2in,angle=0]{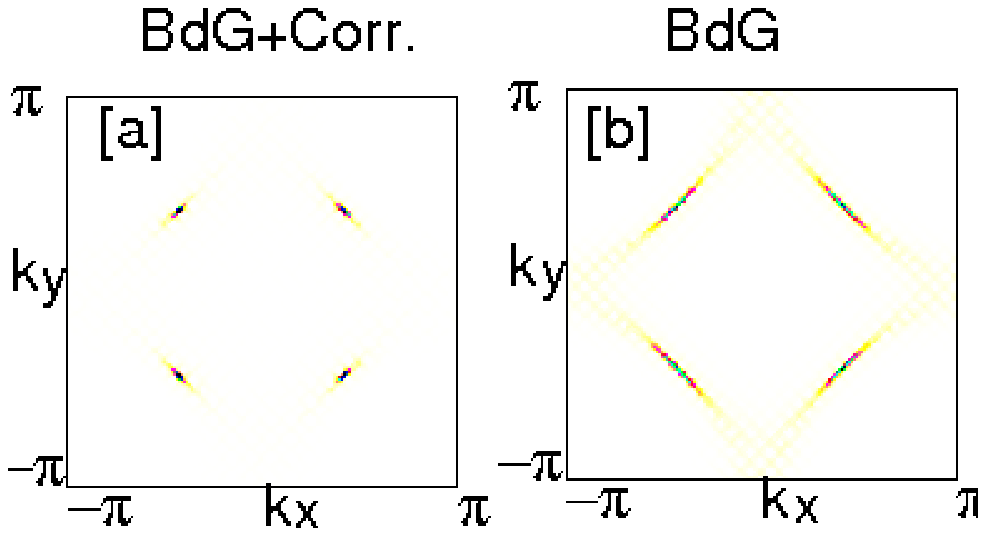}
\vspace{-1.7cm}
\includegraphics[width=1.2in,angle=-90]{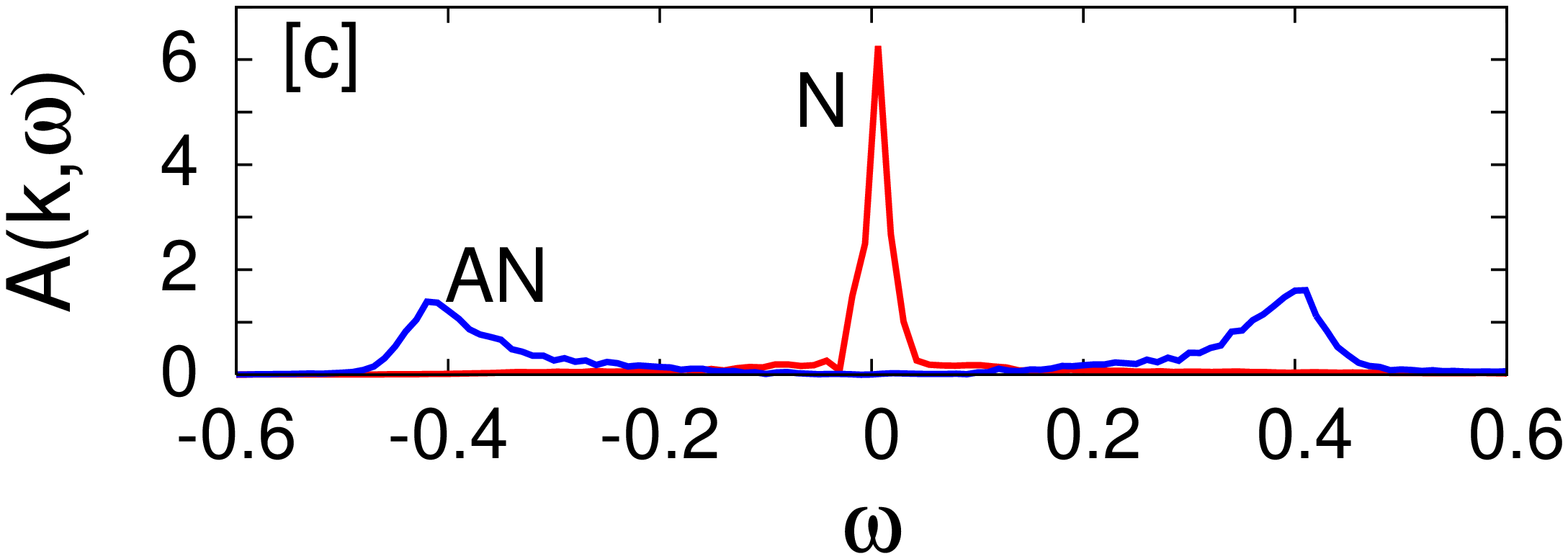}
\vspace{1.3cm}
\caption{
Top: Low energy spectral function $A({\bf k},\omega)$ for $|\omega|\le 0.02t$ for $200\times 200$ system
with $n_{imp}=0.2$. The results with correlations (panel (a)) show only a slight extension around the nodes, while the plain BdG
results (panel (b)) show significant extension into ``Fermi arcs''. (c) $A({\bf k},\omega)$ as a function of $\omega$ for $n_{imp}=0.2$ in the correlated system 
for two momenta: the node (in red) and the antinode (in blue).}
\label{akw}
\end{center}
\vskip-6mm
\end{figure}

\bigskip

\noindent {\bf ARPES spectral function:}
Where do the the low-energy excitations seen in the DOS of Fig.~\ref{dos} come from? To address this question we
found it useful to compute the ARPES intensity, which in a disordered system is given by
the spatially averaged spectral function: 
$A({\bf k},\omega) = \sum_{{\bf r}} \exp(-i{\bf k}\cdot{\bf r}) \left\langle A({\bf r},{\bf R};\omega) \right\rangle_{\bf R}$.
In Fig.~\ref{akw} we plot the near-zero energy intensity $A(\bf {k},\omega)$ and see that the four point nodes of a clean d-wave SC
are extended into ``arcs'' due to disorder \cite{arcs}. 
The gapless excitations at the ${\bf k}$-points on these ``arcs'' contribute to the non-zero DOS. 
A similar effect has been previously found in a T-matrix calculation~\cite{haas} which ignores correlations. 
The surprising new feature of our results is that the extension of the ``arcs'' is much less when 
correlations are taken into account; compare Figs.~\ref{akw}(a) and (b). 
This is yet another manifestation of the fact that disorder effects are suppressed in the presence of strong interactions. 

The calculated ARPES intensities also permits us to compare the behavior of the gapless nodal excitations with the 
maximum gap antinodal states. The coherent piece of the $T=0$ $A({\bf k},\omega)$ in a clean SC consists of 
delta-functions at $\omega=\pm E_{\bf k}$ even in a strongly correlated system \cite{sumrules}. 
In the disordered SC these will be broadened by impurity scattering. We show in Fig.~\ref{akw}(c)
the $\omega$-dependence of $A({\bf k},\omega)$ at two ${\bf k}$-points: the node (N), along the zone diagonal. which is gapless
and the antinode (AN), at the zone edge, which has the maximum of the d-wave gap.

We see from in Fig.~\ref{akw}(c) that the nodal spectrum is much sharper compared to the antinodal one.
A simple golden rule estimate of the impurity scattering rate suggests that the linewidths should scale
like the DOS at the excitation energy. Since we have already seen that the impurity-induced DOS in the correlated system is greatly suppressed 
relative the simple BdG result at low energies, it is self-consistent that the gapless nodal states are much less affected by disorder than the 
gapped antinodal ones.

Our results on the nodal-antinodal dichotomy, obtained by including the effects of strong correlations and inhomogeneity, 
capture the essential features of both ARPES~\cite{mohit_arpes,zhou,shen} and STM~\cite{mcelroy} experiments on cuprates. 
In STM experiments on BISCCO~\cite{mcelroy}, a clear distinction is seen in the behavior of the local DOS for $\omega\approx 0 $ and 
for $\omega$ near the gap. The very low energy spectra are homogeneous and do not vary from one region of the sample to another. 
But the spectra near the gap are very heterogeneous, and thus sensitive to disorder. ARPES experiments~\cite{zhou,shen} 
clearly see the greater sensitivity of the antinodal states to underdoping compared with the nodal states and this has
been attributed to the appearance of competing order at low doping.
Our results indicate that disorder also has the same effect with antinodal excitations strongly affected and the nodal ones relatively robust.

\bigskip

\noindent{\bf Conclusions:}
The calculations presented here focus on the interplay of strong correlations and disorder effects in superconductors, both of which are crucial elements of modeling high $T_c$ cuprates. 
We have shown that correlations suppress the effects of weak impurity scattering leading to 
a shorter healing length around an impurity, a reduction of the impurity-induced low-energy DOS and,
most significantly, robust gapless quasiparticles. We have found that the nodes themselves are much less sensitive to disorder
in a correlated system; in the absence of correlations the nodes expand into ``arcs'' of gapless excitations.
The robust nodal quasiparticles lead to a `V' in the density of states which is highly insensitive to disorder. 
We also found that the large gap antinodal excitations are much more affected by disorder than the gapless nodal ones.
It would be very interesting to 
extend our results to other types of disorder -- such as unitary scatterers \cite{balatsky} and to hopping and superexchnage
disorder \cite{nunner} -- in the presence of strong correlations.

We thank R. Sensarma for discussions.


\begin{thebibliography}{999}
\bibitem{noteU0} We introduce a small onsite repulsion term explicitly $U_0 \sum_i n_{i\da}n_{i\ua}$ with $U_0=t$ which suppresses s-wave pairing and keeps $n(i)\le 1$ on every site.
\bibitem{periodic_repetition}
To generate a denser spectrum we use a periodic repetition scheme with
an $M^2$ array of $L^2$ disordered cells so that the effective size is $N\times N$ with $N=ML$. 
Typical values used are $L=20$ and $M=10$.
\bibitem{incoherent}
The higher energy incoherent spectral weight, not accessible to the Gutzwiller approximation, 
can be constrained by exact sum rules and leads to p-h asymmetry \cite{sumrules}.
\end{thebibliography}
\end{document}